\documentclass[prd,aps,preprint,epsfig,floats]{revtex4}

\newcommand{\beq}{\begin{equation}}
\newcommand{\eeq}{\end{equation}}
\newcommand{\beqn}{\begin{equation*}}
\newcommand{\eeqn}{\end{equation*}}
\newcommand{\bea}{\begin{eqnarray}}
\newcommand{\eea}{\end{eqnarray}}
\newcommand{\bean}{\begin{eqnarray*}}
\newcommand{\eean}{\end{eqnarray*}}

\usepackage{graphicx}

\begin{document}

\draft

\preprint{\vbox{ \hbox{UMD-PP-05-012}}}

\title{\Large\bf $\theta_{13}$ as a Probe of $\mu\leftrightarrow \tau$
symmetry for leptons }
\author{\bf  R.N. Mohapatra  }

\affiliation{ Department of Physics, University of Maryland, College Park,
MD 20742, USA}

\date{August, 2004}

\begin{abstract}
Many experiments are being planned to measure the neutrino mixing
parameter $\theta_{13}$ using reactor as well as accelerator neutrino
beams. In this note, the theoretical significance of a high
precision measurement of this parameter is discussed. It is emphasized
 that it will provide crucial information about different ways to
understand the origin of large atmospheric neutrino mixing and move
us closer towards determining the neutrino mass matrix. For
instance if exact $\mu\leftrightarrow \tau$ symmetry in the
neutrino mass matrix is assumed to be the reason for maximal
$\nu_\mu-\nu_\tau$ mixing, one gets $\theta_{13}=0$. Whether
$\theta_{13}\simeq \sqrt{\Delta m^2_{\odot}/\Delta m^2_A}$ or
$\theta_{13}\simeq \Delta m^2_{\odot}/\Delta m^2_A$ can provide
information about the way the $\mu\leftrightarrow \tau$ symmetry
breaking manifests in the case of normal hierarchy. We also discuss the
same question for inverted hierarchy as well as possible gauge theories
with this symmetry.
 \end{abstract}

\maketitle

\section{Introduction}
Neutrino physics is poised on the brink of an exciting set of
experiments that could elevate our knowledge of neutrino masses
and mixings to the same level as that of quarks and charged
leptons. At the same time, they are also likely to provide
important information about physics beyond the standard model.
The most crucial experiments in this regard are: (i) searches for
neutrinoless double beta decay which will confirm whether
neutrinos are Dirac or Majorana fermions; (ii) sign of the
atmospheric mass difference which will determine whether the mass
hierarchy is normal or inverted and (iii) the magnitude of the
unknown angle $\theta_{13}$, which will complete our knowledge of
mixings.

In this article, I discuss the impact of a high precision search
for $\theta_{13}$ assuming that neutrinos are Majorana fermions. There are
several experimental proposals for
such searches e.g. Ref.\cite{reactor,bnl}. Some of these experiments are
also likely to yield a more precise value of the atmospheric neutrino
mixing angle.
The value of $\theta_{13}$ in addition to providing a complete picture of
neutrino mixings, could be a signal of the underlying
physics responsible for lepton mixings and as such could be an
important clue to physics beyond the standard model\cite{review}. As is
argued in this paper, value of $\theta_{13}$ in conjunction with a
high precision measurement of the maximality of the atmospheric mixing
angle $\theta_A\equiv \theta_{23}$ could indeed be a very useful way to
determine the complete neutrino mass matrix for the case of a normal
hierarchical spectrum for neutrinos.

To begin the discussion, let us note that the PNMS mixings arise
from the lepton mass Lagrangian as follows:
\begin{eqnarray}
{\cal L}_m~=~\nu^T_\alpha C^{-1}{\cal M}_{\nu,\alpha\beta}\nu +
\bar{e}_{\alpha,L}M^e_{\alpha \beta}e_R + h.c.
\end{eqnarray}
Diagonalizing the mass matrices by the transformations
$U^T_\nu{\cal M}_\nu U_\nu={\cal M}^\nu_{diag}$ and $U^{\dagger}_\ell
M^eV~ =M^e_{diag}$, one defines
$U_{PMNS}~=~U^{\dagger}_\ell U_\nu$. Clearly, any symmetry in the
lepton mass matrices is likely to manifest itself in the
$U_{PMNS}$ elements, at least in the basis where the charged leptons
are mass eigenstates. We will parameterize $U_{PMNS}$ as follows:
\begin{eqnarray}
 U_{PMNS}=\pmatrix{c_{12}c_{13} & s_{12}c_{13} &
s_{13} e^{-i\delta} \cr
-s_{12}c_{23}-c_{12}s_{23}s_{13}e^{i\delta}
&c_{12}c_{23}-s_{12}s_{23}s_{13}e^{i\delta} & s_{23}c_{13}\cr
s_{12}s_{23}-c_{12}c_{23}s_{13}e^{i\delta}
&c_{12}s_{23}-c_{12}c_{23}s_{13}e^{i\delta} &
 c_{23}c_{13}}K
\end{eqnarray}
where $K~=~diag(1, e^{i\phi_1},e^{i\phi_2})$.

To see how the symmetry of the mass matrix appears in the mixing
matrix, let us consider the case of only two neutrino generations
i.e. that of $\mu$ and $\tau$. Experiments indicate that the
atmospheric mixing angle is very nearly maximal i.e. $\theta_A~=~
\pi/4$.  Working in the basis where the charged lepton mass
matrix is diagonal, it is obvious that the nautrino Majorana mass
matrix that gives maximal mixing is:
\begin{eqnarray}
{\cal M}^{(2)}_\nu~=~\pmatrix{a & b\cr b & a}.
\end{eqnarray}
Furthermore the fact that solar neutrino mass difference square
$\Delta m^2_\odot \ll \Delta m^2_A$ and allowing for small
departures from the maximal atmospheric angle, we can write
\begin{eqnarray}
{\cal M}^{(2)}_\nu~=~\frac{\sqrt{\Delta
m^2_A}}{2}\pmatrix{1+a\epsilon & 1\cr 1 & 1+\epsilon}
\end{eqnarray}
where $a$ is a parameter of order one and $\epsilon \ll 1$. For the case
of normal hierarchy we have $\sqrt{\Delta
m^2_\odot/\Delta m^2_A}\simeq \frac{1}{4}(1+a)\epsilon$. The
atmospheric mixing angle is given by $\theta_A\simeq
\frac{\pi}{4}-\frac{\epsilon(1-a)}{4}$. It is clear if $a=1$, the
neutrino mass matrix has symmetry $\nu_\mu\leftrightarrow
\nu_\tau$ and $\theta_A=\pi/4$. Thus departures from this symmetries
remain imprinted in the values of the mixing angles.

\section{Exact $\nu_\mu\leftrightarrow \nu_\tau$ symmetry and
$\theta_{13}=0$}

Let us now extend the above considerations to
the case of three generations. First point to note is that in
the zeroth order, clearly unrealistic, approximation,
 maximal atmospheric mixing can arise from two kinds of
neutrino mass matrices:

\noindent{\it Case (i)}:
\begin{eqnarray}
{\cal M}_\nu~=~\frac{\sqrt{\Delta m^2_A}}{2}\pmatrix{0&0&0\cr 0 &
1 & 1 \cr 0 & 1 & 1}
\end{eqnarray}
This is the case of normal hierarchy.

\noindent{\it Case (ii)}:
\begin{eqnarray}
{\cal M}_\nu~=~\frac{\sqrt{\Delta m^2_A}}{2}\pmatrix{0&1&1\cr 1 &
0 & 0 \cr 1 & 0 & 0}
\end{eqnarray}
This is the case of inverted hierarchy. Both these mass matrices are
 invariant under $\nu_\mu\leftrightarrow \nu_\tau$
symmetry. Furthermore, the second case has the additional
symmetry : $L_e-L_\mu-L_\tau$\cite{emutau}. In both cases of
course one has $\Delta m^2_\odot =0$; $\theta_{23}=\pi/4$ and
$\theta_{13}=0$. For the matrix in Eq. (6), one also has in addition
$\theta_{12}=\pi/4$.

In order to depart from this unrealistic zeroth order case to the
more realistic case and to see how the various mixing angles are
affected, let us first ask the question as to whether one can have mass
matrices invariant under  $\nu_\mu\leftrightarrow \nu_\tau$
symmetry while giving  $\Delta m^2_\odot \neq 0$ and $\theta_{12}
< \pi/4$. The answer to this question is ``yes''. An example of such a
mass matrix is:
\begin{eqnarray}
{\cal M}_\nu~=~\frac{\sqrt{\Delta m^2_A}}{2}\pmatrix{c\epsilon
&d\epsilon &d\epsilon\cr d\epsilon & 1+\epsilon & -1 \cr
d\epsilon & -1 & 1+\epsilon}
\end{eqnarray}
Mass matrices of this type have been considered in \cite{ma}.  A
mass matrix with $\Delta m^2_\odot \neq 0$ but $\theta_{12} =
\pi/4$ was discussed early on from considerations of
$\nu_\mu\leftrightarrow \nu_\tau$ symmetry in
\cite{nuss}. Both these \cite{lam,nuss} $\nu_\mu\leftrightarrow
\nu_\tau$ symmetric neutrino mass matrices lead to
$\theta_{13}=0$.

For this mass matrix, we have
\begin{eqnarray}
\epsilon~=~4\sqrt{\frac{\Delta m^2_\odot}{\Delta m^2_A}}\frac{1}{[(c+1)+
\sqrt{(c-1)^2+8d^2}]}\\ \nonumber 
tan 2\theta_\odot~\simeq \frac{2\sqrt{2}d}{1-c}\\ \nonumber
\theta_{23}~=~\frac{\pi}{4}; \theta_{13}~=~0.
\end{eqnarray}
Thus two of the three parameters of this matrix are determined by
already existing data and if $\theta_{13}$ is found to be smaller
than the limit expected in many forthcoming experiments and it is
found that $\Delta m^2_{31} >0$, then there would be a strong
case for the matrix in Eq. (7) as the mass matrix for the neutrinos (in
the basis
where the charged leptons are mass eigenstates)  as well as for an
underlying $\nu_\mu\leftrightarrow \nu_\tau$ symmetry. A test
of this mass
matrix would be a value of $\theta_{23}~=~\pi/4$. Since
neutrinoless double beta decay can in principle determine the
parameter $c$, one can determine all the parameters of this model.
This would clearly be a major step forward in probing physics beyond the
standard model.

We discuss the case of inverted mass hierarchy in
subsequent section using the results in Ref.\cite{goh}.

\section{Departures from $\nu_\mu\leftrightarrow \nu_\tau$ symmetry and
expectations for $\theta_{13}$}

We now consider perturbations around the symmetric limit for the
normal hierarchy case and discuss its consequences. Many
discussions of such cases exist in the literature\cite{others},
(though not necessarily in the context of
$\nu_\mu\leftrightarrow \nu_\tau$ symmetry). We motivate our discussion
from the angle of this symmetry. We mostly discuss
the case without CP violation and in the end of this section, comment on a
case with CP violation.

The most general CP conserving perturbation of the neutrino mass
matrix around the $\nu_\mu\leftrightarrow \nu_\tau$ symmetric
limit that maintains the hierarchy $\Delta m^2_\odot \ll \Delta
m^2_A$ and near maximal atmospheric mixing is:
\begin{eqnarray}
{\cal M}_\nu~=~\frac{\sqrt{\Delta m^2_A}}{2}\pmatrix{c\epsilon
&d\epsilon &b\epsilon\cr d\epsilon & 1+a\epsilon & -1 \cr
b\epsilon & -1 & 1+\epsilon}
\end{eqnarray}
The parameters characterizing the departures from symmetry limit
are: $b\neq d$ and $a\neq 1$. Two characteristic predictions
appear depending on the way the symmetry breaking appears in the mass
matrix.

\noindent{\it Case (i): $a=1, b\neq d$}

 In this case, we diagonalize the mass matrix for the case when $c\ll 1$
and ignoring terms of order $(b-d)/(b+d)$ in $\epsilon$ but keeping them
in $\theta_{13}$. (Keeping these terms in $\epsilon$ gives a somewhat
complicated expression and since we are interested in qualitative
predictions, we do not include these corrections). We find that
\begin{eqnarray}
\epsilon \simeq \frac{4}{1+\sqrt{1+8d^2}}\sqrt{\frac{\Delta
m^2_\odot}{\Delta
m^2_A}}\\
\nonumber
 \theta_{13}\simeq (b-d)\sqrt{\frac{\Delta m^2_\odot}{
\Delta m^2_A}}\\ \nonumber tan 2\theta_{\odot}\simeq
\frac{2(b+d)}{1-c} \\ \nonumber
m_{\beta\beta}\simeq c\epsilon
\end{eqnarray}
 Using present
data, in this case one would expect $\theta_{13}$ slightly below its
present upper limit (say around 0.15 or so). The
predictions in models where atmospheric neutrino mixing arises
from some dynamical mechanism\cite{dyn} are also similar.
 The difference between this approximate
$\mu\leftrightarrow \tau$ symmetry case and the ``dynamical'' case
is that the atmospheric mixing angle in the symmetry case being discussed
here is very close to maximal with departure from maximality being of
order $\frac{\Delta m^2_\odot}{\Delta m^2_A}$ which is a few per cent
(of order $\leq 4^0$) whereas in the dynamical case, this departure can
 be larger (of order $\sim 8^0$ or so). The prediction for neutrinoless
double beta decay in this case is beyond the range of accessibility of the
next round of searches for double beta decay\cite{db}.

 The physical meaning of this case is
that while $\nu_\mu\leftrightarrow \nu_\tau$ symmetry is exact in
the $\nu_\mu-\nu_\tau$ sector, it is broken in their mixing with
$\nu_e$. We will call this e-sector breaking. Unless this breaking is
constrained by extra symmetries,
one would expect a large $\theta_{13}$, as noted.

\noindent{\it Case (ii): $a\neq 1, b= d$}

In this case, we get
\begin{eqnarray}
\epsilon \simeq
\frac{4}{[c+(1+a)/2]+\sqrt{[c-(1+a)/2]^2+8d^2}}\sqrt{\frac{\Delta
m^2_\odot}{\Delta m^2_A}}\\
\nonumber \theta_{13}\simeq
\frac{1}{4\sqrt{2}}\epsilon^2d(1-a)
\end{eqnarray}
In this case there is a departure from maximality of the
atmospheric mixing angle given by the following equation:
\begin{eqnarray}
\theta_A\simeq \frac{\pi}{4}-\epsilon\frac{1-a}{4}
\end{eqnarray}
Thus, the  expectation for $\theta_{13}$ for this way of symmetry breaking
is around $\theta_{13}\approx
0.03$. The smallness of $\theta_{13}$ here compared to the
previous case can be understood as follows: the
$\nu_\mu\leftrightarrow \nu_\tau$ symmetry is broken in the only
in the $\nu_\mu-\nu_\tau$ sector of the mass matrix and not in
the mixing with $\nu_e$. As a result, to leading order in
$\epsilon\approx\sqrt{ \frac{\Delta m^2_\odot}{\Delta
m^2_A}}$, there is no contribution to $\theta_{13}$ and it arises
only to order $\epsilon^2$.  Also as noted above, the departure from
maximality of the atmospheric mixing angle in this case can be significant
($\sim 8-10\%$).

\noindent{\it Case (iii): $a=1; |b|=|d|$} An interesting way to
break $\nu_\mu\leftrightarrow \nu_\tau$ is to maintain $a=1$ so
that symmetry breaking is in the mixing with $\nu_e$; but choose
$b = d^*$\cite{grimus}. In this case, one has $\theta_{13}=2Im
b\sqrt{\frac{\Delta m^2_\odot}{ \Delta m^2_A}}$ and one has the
Dirac phase at its maximal value of $\pi/2$.

In the table below, we summarize our results:

\begin{center}
\begin{tabular}{|c||c||c|}\hline
symmetry breaking & $\theta_{13}$ & $\theta_{23}-\pi/4$ \\ \hline
None & 0 & 0 \\
$\mu-\tau$ sector only & $\sim \Delta m^2_\odot/\Delta m^2_A$ & $\leq
8^0$\\
e-sector only &  $\sim \sqrt{\Delta m^2_\odot/\Delta m^2_A}$ &
$\leq 4^0$\\
dynamical &  $\sim\sqrt{ \Delta m^2_\odot/\Delta m^2_A}$ & $\leq 8^0$ \\
\hline
\end{tabular}
\end{center}

\noindent{\bf Table caption:} This table gives the predictions for
$\theta_{13}$ and $\theta_A$ for different ways of
$\mu\leftrightarrow \tau$ symmetry breaking. Note that what we mean by
e-sector only is that $\mu\leftrightarrow \tau$ symmetry is broken in the
$e-\mu$ and $e-\tau$ elements of the Majorana neutrino mass matrix.
Similarly, when we say $\mu-\tau$ sector, we mean the symmetry is broken
in $\mu-\tau$ subsector of the neutrino mass matrix.

\section{Departures from $\nu_\mu\leftrightarrow \nu_\tau$ symmetry: the
inverted hierarchy case:}
 The case of inverted hierarchy has been discussed in great detail in
\cite{goh} (although connection to $\mu\leftrightarrow \tau$
symmetry was not discussed). Here I summarize the discussion in the
language of $\mu\leftrightarrow \tau$ symmetry.

 The most general mass matrix in this
case is:
\begin{eqnarray}
{\cal M}_\nu=\sqrt{\Delta m^2_A}~\left(\begin{array}{ccc} z &
c & s\\ c & y & d\\ s & d & x\end{array}\right).
\end{eqnarray}
where $c$ and $s$ stand for $cos$ and $sin$ of $ \theta_{23}$ and $x, y,
z, d \ll 1$. In the perturbative
approximation, we find the following sumrules
involving the neutrino observables and the elements of the neutrino mass
matrix.  It follows from this matrix that
\begin{eqnarray}
\sin^22\theta_{\odot}~=~1-(\frac{\triangle m_\odot^2}{4\triangle
m_A^2}-z)^2~+~O(\delta^3) \cr
 \frac{\triangle m_\odot^2}{\triangle
m_A^2}~=~2(z+\vec{v}\cdot\vec{x})~+~O(\delta^2)\cr
U_{e3}~=~\vec{A}\cdot(\vec{v}\times\vec{x})~+~O(\delta^3)\cr
\end{eqnarray}
where
$\vec{v}=(\cos^2\theta,\sin^2\theta,\sqrt{2}\sin\theta\cos\theta)$,
$\vec{x}=(x,y,\sqrt{2}d)$ and
$\vec{A}~=~\frac{1}{\sqrt{2}}(1,1,0)$. $\delta$. Now we can
discuss the exact $\nu_\mu\leftrightarrow \nu_\tau$ limit and
departures from it. The exact symmetry limit occurs when we have
$c=s=\frac{1}{\sqrt{2}}$ (maximal atmospheric mixing angle) and
$x=y$. It is clear from above that $\theta_{13}=0$ in this limit.
Therefore, a nonvanishing $\theta_{13}$ is related to breakdown of
this symmetry as in the case of normal hierarchy.

It is clear from this way of parameterizing the mass matrix that
the current best fits for the large mixing angle solution to the
solar neutrino observations\cite{fogli} require $z \geq 0.3$ or
so. This translates into a lower limit on $m_{\beta\beta}\geq 15$
meV\cite{idb}. Similar to the case of normal hierarchy case,
there are two broken symmetry situations.

\noindent{\it Case (i): $c=s=\frac{1}{\sqrt{2}}$; $x\neq y$}:
In this case, we have
\begin{eqnarray}
\theta_{13}~=~\frac{x-y}{2}\cr
\frac{\triangle m_\odot^2}{\triangle
m_A^2}~=~2(x+y+z+d)
\end{eqnarray}
In this case, $\theta_{13}$ could be quite large. It is worth
noting that in this case even though smallness of $\frac{\triangle
m_\odot^2}{\triangle m_A^2}$ implies that there must be
cancellations among the parameters $x,y,z$ and $d$, it does not
put any constraint on how large $\theta_{13}$ can be.

\noindent{\it Case (ii): $c\neq s$;  $x~=~ y$}
In this case we find
\begin{eqnarray}
\theta_{13}~\simeq -dcos 2\theta_A
\end{eqnarray}
In this case, there is a close connection between the value of
$\theta_{13}$ and departure from maximality of $\theta_A$.

It is clear that the expectations for $\theta_{13}$ for the inverted
hierarchy are very different from the normal hierarchy case. Specially
missing in this case is the close connection between $\theta_{13}$ and
the ratio $\frac{\triangle m_\odot^2}{\triangle
m_A^2}$. The reason for this is that the value of $sin^22\theta_\odot$
required by the present solar and KamLand data requires the $m_{ee}$ term
in the neutrino mass matrix to be large in the case of inverted hierarchy.
This therefore enters as a new parameter in the $\Delta m^2_\odot$ unlike
the case of normal hierarchy.

\section{Possible gauge theory of broken $\nu_\mu\leftrightarrow
\nu_\tau$}

So far the discussion has focussed on the testability of
$\nu_\mu\leftrightarrow \nu_\tau$ symmetry in the neutrino
Majorana mass matrix. In this section we would like to address
its implications for physics beyond the standard model. We would
like to seek plausible gauge models that lead to this symmetry. We will
focus only on the normal hierarchy case since the case of inverted
hierarchy has been studied extensively in the literature\cite{emutau}.

The first clear obstacle one must overcome is that the neutrinos
are part of the $SU(2)_L$ doublet that contains the charged leptons
$(e,\mu,\tau)$ and there is  no apparent $\mu\leftrightarrow \tau$
symmetry in the charged lepton masses. However, in the limit of
$m_\mu=m_\tau$, one can have such a symmetry implying that in the charged
lepton sector, there must clearly be a mechanism to break the
symmetry by a large amount without affecting the neutrino. We explore
below how such a symmetry
can emerge in gauge theories and in particular, how it can be
broken in a consistent manner. The goal is a modest one of simply
trying to give an existence proof. The main point is that if one
cannot even construct a consistent model within the loose
framework of arbitrary fine tuning, the symmetry has a less
chance of being meaningful in reality. In our case it turns out
that in addition to assuming a spontaneously broken
$\mu\leftrightarrow \tau$ symmetry, if one assumes a $Z_4$
symmetry, then there are several models that one can construct
that realize the mass matrix in Eq. (7) without conflicting with charged
lepton spectrum with rather mild assumptions. We only discuss the
symmetric limit. One can easily
extend them to include small breaking effects e.g. by adding higher
dimensional terms to the Lagrangian.

We first show that for $\theta_{13}$ to vanish, the $\mu\leftrightarrow 
\tau$
symmetry must be in the left handed neutrino sector; in other words, if we
had the permutation symmetry only in the RH neutrino sector, it does not
lead to a vanishing $\theta_{13}$. We then present two
models one with right handed neutrinos and one without them where
$\mu\leftrightarrow \tau$ symmetry is imposed both in the left and right
handed sector and show that it leads to vanishing $\theta_{13}$ in the 
symmetry limit. In the first
case we will use the conventional seesaw mechanism and in the second one,
we will use a triplet dominated type II seesaw\cite{type2}.

\subsection{ Model I:}
 We now consider models where $\mu\leftrightarrow \tau$ applies both in
the left and right handed neutrino sector. We use the
standard model gauge group with supersymmetry and standard
assignment of matter superfields\cite{book} but with three pairs
of Higgs doublets ($H_u,H_d$). We impose on the model an
$S_2\times Z_4$ symmetry. The multiplets $(L_\mu,L_\tau)$,
$(\mu^c,\tau^c)$, $(N^c_\mu,N^c_\tau)$, $(H_{d,1}, H_{d,2})$
transform into each other under $S_2$ symmetry and $H_{u,i}$ (i=1,2,3)
and the rest of the fields transform as singlets. Under $Z_4$, we assign
$(\mu^c, H_{d,2})$ to
transform as $i(\mu^c, H_{d,2})$ whereas $(\tau^c, H_{d,1})$ go to
$-i(\tau^c, H_{d,1})$. Rest of the fields are invariant. First
point to note is that, the right handed neutrino mass matrix
invariant under this has the form
\begin{eqnarray}
M_R~=~\pmatrix{M_{11} & M_{12} & M_{12}\cr M_{12} & M_{22} &
M_{23}\cr M_{12} & M_{23} & M_{22}}
\end{eqnarray}

The Dirac mass matrix for the neutrinos also has similar form:
\begin{eqnarray}
m_D~=~\pmatrix{m_{11} & m_{12} & m_{12}\cr m_{12} & m_{22} &
m_{23}\cr m_{12} & m_{23} & m_{22}}
\end{eqnarray}
It is clear that the neutrino mass matrix obtained from the above
two equations after type I seesaw  has the form which is as in
Eq.(7) which is $\mu\leftrightarrow \tau$ invariant.

To complete the discussion of model I, note that the charged lepton
mass matrices arise from the superpotential:
 \begin{eqnarray}
 W~=~h_1 (L_\mu H_{d,1}\mu^c+L_\tau H_{d.2}\tau^c) + h_e L_e
 H_{d,3}e^c+h_3(L_\tau H_{d1}\mu^c+L_\mu H_{d2}\tau^c)
 \end{eqnarray}
 Now if we set $h_3=0$ and suppose that we break the
 $\mu\leftrightarrow \tau$ symmetry by the soft $H_{d,1,2}$ mass
 terms, then $H_{d,1,2}$ will have different and arbitrary vevs. As a
result, we can get correct values for all the charged lepton masses.

\subsection{ Model II without right handed neutrinos}

This model is
very similar to the model above except that there are no right
handed neutrinos- instead there are Higgs triplets $\Delta_L$ with
standard model hypercharge +2 so that couplings of type
$LL\Delta_L$ are allowed. The $\Delta_L$ is given a mass term $M$
which is of order of the $10^{14}$ GeV, so that the vev of
$\Delta_L$ is suppressed due to the term $\Delta_L H_dH_d$ to be
$v^2_{wk}/M\simeq 10^{-1}$ eV, which
can give neutrino masses of the right order. As in the first
case, we require the model to be invariant under $S_2\times Z_4$
symmetry with assignments as in the previous case. The fields in
$\Delta_L\oplus \bar{\Delta}_L$ pair are invariant under it. The
neutrino masses come from the superpotential
\begin{eqnarray}
f_1(L_\mu +L_\tau)((L_\mu +L_\tau))\Delta_L +f_2(L_\mu
-L_\tau)((L_\mu -L_\tau))\Delta_L +(L_\mu+L_\tau)L_e\Delta_L
+L_eL_e\Delta_L
\end{eqnarray}
Again this leads to a neutrino mass matrix invariant under
$\mu\leftrightarrow \tau$ symmetry. The charged lepton masses
arise in exactly the same way as in the model I.

There are also other models in the literature with similar
properties (e.g. see ref.\cite{fr}) also. It would therefore seem that
considering $\mu\leftrightarrow \tau$ symmetry for leptons, despite its
strong breaking in the charged lepton sector is quite a meaningful and
useful way to obtain information about physics beyond the standard model
from neutrinos.

Let us make a few comments on the models described above. Note
that we have not incorporated any breaking of $\mu\leftrightarrow
\tau$ into the model. There could many many sources for such
breakings: for example, there could be higher dimensional
operators that involve $H_{d,1,2}$ that can break this symmetry.
There could also be other effects such as radiative corrections
from charged lepton Yukawa couplings that give mass to tau lepton
and the muon etc.

\section{Summary and conclusion}
In this brief note, it is pointed out that the measurement of the
neutrino mixing angle $\theta_{13}$ in conjunction with a measurement
of the departure from maximality of the atmospheric mixing angle can be a
very powerful way to probe
any possible $\nu_\mu\leftrightarrow \nu_\tau$ symmetry present in the
neutrino mass matrix. In Table I, the expectations for $\theta_{13}$ and
different cases (with and without approximate  $\nu_\mu\leftrightarrow
\nu_\tau$ symmetry) are presented for the case of normal hierarchy and can
be used as a way to specify the mass matrix. We also have discussed the
case of inverted mass hierarchy and pointed out the implications of broken
$\mu\leftrightarrow \tau$ symmetry.

Evidence for any approximate  $\nu_\mu\leftrightarrow \nu_\tau$ symmetry
will clearly be a significant indicator of which way to proceed as
we probe physics beyond the standard
model. For instance, such a symmetry is highly nontrivial to obtain within
the framework of grand unification and pint to alternative directions,
which will be a useful information.

 We must emphasize that all our considerations are based on
the assumption that there are no extra sterile neutrinos mixing with the
three known active ones. Claerly therefore any evidence for sterile
neutrinos will require a re-evaluation of the conclusions stated in the
paper.

This work is supported by the National Science Foundation grant
no. Phy-0354401. I like to thank Salah Nasri and the anonymous referee
for suggesting improvements.

Note added: The results of this paper were presented at the APS
neutrino study wrap-up meeting at Snowmass in June, 2004 (
http://www.neutrinooscillation.org/studyaps/apsfinalprogram.html)
and at SLAC Summer Institute on August 7, 2004
(http://www-conf.slac.stanford.edu/ssi/2004)
After this work was completed, a paper by W. Grimus et al
(hep-ph/0408123) with similar conclusions appeared in the arXivs.

\end{document}